\documentclass[aps,prl,twocolumn,superscriptaddress,preprintnumbers,nofootinbib]{revtex4-1}

\usepackage{times}

\begin{document}

\title{The forest as a neutrino detector}

\author{S.~Prohira}

\affiliation{University of Kansas, Lawrence, KS 66045, USA}

\begin{abstract}
    The primary challenge in detecting ultrahigh energy (UHE) neutrinos with energies exceeding $10^{16}$\,eV is to instrument a large enough volume to detect the extremely low flux, which falls as $\sim E^{-2}$. We explore in this article the feasibility of using the forest as a detector. Trees have been shown to be efficient broadband antennas, and may, without damage to the tree, be instrumented with a minimum of apparatus. A large scale array of such trees may be the key to achieving the requisite target volumes for UHE neutrino astronomy. 
\end{abstract}

\maketitle

{\bf Introduction.---} Detection of neutrinos in the ultrahigh energy (UHE) range relies upon instrumenting large target volumes with a minimum of apparatus. Current and planned detectors~\cite{IceCube:2002eys, BAIKAL:1997iok, KM3Net:2016zxf, ARA:2014fyf, ANITA:2008mzi, ARIANNA:2014fsk, Nam:2016cib, GRAND:2018iaj, Otte:2018uxj, RNO-G:2020rmc, PUEO:2020bnn, POEMMA:2020ykm, Wissel:2020sec, Romero-Wolf:2020pzh, P-ONE:2020ljt, RadarEchoTelescope:2021rca} use various technologies to instrument target volumes of water, air, and ice to monitor for neutrino interactions. Those techniques that monitor the air, such as GRAND, BEACON, TRINITY, TAMBO, or TAROGE-M, seek detection of the in air decay of tau leptons emerging from a tau neutrino interaction in the Earth. These experiments have exceptional projected sensitivity~\cite{Ackermann:2022rqc} for full scale arrays at various energies, and many have successfully deployed demonstrator instruments for proof of concept~\cite{Southall:2022yil,Martineau-Huynh:2019bgk, Otte:2023osf,TAROGE:2022soh}. 

The primary challenge for scalability from demonstrator to full size array is to find suitable locations, particularly for those methods that detect tau decays via coherent radio emission. Antennas require a clear line of sight to the horizon (specifically, the $\tau$ exit point just below the horizon) and require distance from civilization to minimize interference from anthropogenic sources. Other challenges to scalability are practical, including the difficulty of instrumenting large volumes with tens to hundreds of thousands of manufactured antennas (GRAND) or the relative inaccessibility of the deployment site (TAROGE-M). The experiments listed here have achieved success in overcoming these difficulties for prototype instruments, but they persist as challenges to scalability for full sized arrays. 

In this letter we present a simple idea: use the forest as the instrument, with trees as radiofrequency antennas~\cite{treeAntenna,ikrath1975trees, ad735330,ikrath1973utilization, ikrath1972performance, kar2010trees,liao2023performance}. There are three primary advantages of using trees as antennas for a radio UHE $\nu_{\tau}$ detector:
\begin{enumerate}
\item Minimum of apparatus. Trees are sturdy, resilient, and already established in forests, obviating the need to construct and deploy robust manufactured antennas. As we discuss in this letter, the apparatus to connect a tree to readout electronics is simple, and therefore promising for scalability.
\item Coverage. Instrumenting a number of trees on a wooded hill or mountain results in a uniformly instrumented array, without the need to find barren hills to instrument.
    \item Frequency band. A limitation of most detection technology that relies on optical or radio-frequency detection is a limited band for detection, owing to the difficulties of making broadband antennas of a reasonable size. Trees are intrinsically broadband detectors, with existing studies documenting trees as antennas from VLF (3-30\,kHz) up through HF (3-30\,MHz), motivating more study at frequencies greater than 30\,MHz.

\end{enumerate}

Such an instrument could be deployed in any number of forests in any number of locations all around the globe, in collaboration with local communities and governing agencies. In this letter we summarize existing studies and use them to provide an estimate of sensitivity for a forest detector.

{\bf Trees as antennas.---}\label{treeasantenna} Studies beginning in the 20th century demonstrated clearly that trees were efficient broadband antennas.\footnote{Though ref.~\cite{treeAntenna} notes that ``a tree is as good as any man-made aerial, regardless of the size of the extent of the latter, and better in the respect that it brings to the operator's ears far less static interference", this assertion from 1919 about the low noise figure is not quantified.}  Their study was originally military in nature, the idea being that trees could serve as both transmitters and receivers in remote situations. Comparisons of reception performance relative to manufactured antennas in varying weather conditions were performed with systems that inductively couple the tree to a receiver system via a simple toroidal coil~\cite{ikrath1975trees}. Phasing of trees (and other structures) was performed in~\cite{ikrath1973utilization}, indicating that multiple trees may be employed as a phased radio array. In studies of jungle trees~\cite{ikrath1975trees,ikrath1972performance}, the instrumented trees outperformed manufactured antennas for 4.6\,MHz in terms of signal strength and signal to noise ratio, in some cases by up to 20\,dB. Promisingly, recent patents demonstrate active research into utilizing trees as efficient antennas in the VHF (3-300\,MHz) range~\cite{treeantennapatent}.

Tree antennas in these studies were connected to electronics in two ways. The original method, seemingly best suited for MF (300\,KHz-3\,MHz) to HF transmission, is to drive a nail into the tree and run a wire from this nail. A separate lead is connected to a grounding rod, and these are connected to the readout electronics. Later studies employed a toroidal coil wrapped around the tree 1-2\,m above grade, inductively coupling the tree to the readout electronics. Both systems demonstrated improvements over manufactured monopole or dipole antennas of the same length and configuration (when insulated from the tree) or the toroidal coil in free space.

 The effects of foliage on HF and VHF signals using traditional manufactured antennas in forested jungles were reported in~\cite{wait1974workshop, sturgill1967tropical, hicks1969tropical,robertson1969tropical} and UHF signals in~\cite{nelson1980uhf}. The HF/VHF studies indicate that horizontally polarized radiation is less affected by foliage than vertical, and the height of a receiving antenna with respect to the surrounding foliage and the height of the radio signal has an effect on propagation losses. For a receiving antenna with line of sight to a transmitting antenna, intervening foliage has a minimal effect on propagation (consistent with a smooth earth model), which would be the case for a tree antenna in a hillside forest as a receiver for a source at or just below the horizon.

{\bf Neutrino detector.---} We can estimate the sensitivity of a forest detector. Current studies that employ arrays of manufactured antennas give sensitivities based on instrumented area. We can estimate a sensitivity relative to these detailed studies by making reasonable estimates of both the density of instrumentation and the quality of the antennas. 

The average spacing of lodgepole pine trees ({\it Pinus contorta}) in the Rocky Mountains is comparable to the average diameter of the extent of their mature foliage, which can range from 2-5\,m\footnote{According to the author's recollections.}. Pine trees are also less dense in terms of specific leaf area than broadleaf deciduous trees~\cite{richardson2000ecology}, which may result in decreased signal attenuation for less steeply sloped arrays. It is therefore feasible to assume that instrumentation of an array with the exact configuration of that presented in, for example, ref.~\cite{GRAND:2018iaj} is achievable by connecting trees of the desired antenna spacing (tuning of which can set an energy threshold). A relative estimate of overall array sensitivity is then obtained by calculating the relative efficiency of the tree antennas to the manufactured antennas assuming a comparable array geometry. Furthermore, successful phasing of tree antennas was discussed in refs.~\cite{ikrath1975trees,ikrath1973utilization}, indicating that the successful BEACON concept~\cite{Wissel:2020sec} may be employed using these natural antennas to provide directivity.

The impulsive radio produced in tau decays from neutrino interactions is intrinsically broadband, with higher signal strengths and wider geometric acceptance at HF~\cite{Wissel:2020sec}. Studies of tree antennas at 4.6\,MHz showed improvements of 6-20\,dB over a short monopole ``whip'' style antenna tuned to the same frequency, depending on the weather~\cite{ikrath1975trees}. In this study, four transmission cases were tested: whip to whip, tree to whip, whip to tree, and tree to tree. Received signal strength was measured as a function of distance between transmitter and receiver. In all weather, tree to tree transmission performed the best, then whip to tree, then tree to whip, and finally whip to whip. Assuming a representative improvement of $\sim$10\,dB (whip to tree compared to whip to whip in dry weather, Fig.~7 of Ref.~\cite{ikrath1975trees}), we can estimate the HF gain of the tree antenna to be $\geq$4\,dBi.\footnote{The the PRC-74 whip used in Ref.~\cite{ikrath1975trees} has a length of 10\,ft so we estimate its realized gain at 4.6\,MHz as -6\,dBi. This however does not account for the matching network of the unit, which increases this realized gain.} Sensitivities for the lower frequency band in~\cite{Wissel:2020sec} use antenna gains of 1.8\,dBi, indicating that tree antennas may provide comparable $\nu_{\tau}$ sensitivity at these frequencies. We note, however, that 4.6\,MHz is just below the simulated range in~\cite{Wissel:2020sec}, and that the average noise temperature of an antenna pointed to the horizon is greater at 4.6\,MHz relative to VHF ($>$30\,MHz)~\cite{GRAND:2018iaj}, which may offset of the advantage in antenna sensitivity, further motivating study of trees as antennas at VHF frequencies, where radio $\nu_{\tau}$ detectors operate. Preliminary experimentation by the author at VHF indicates promising sensitivity.~\cite{tree_exp_prelim} 
 
The readout electronics for the forest detector would be largely identical to those of any other large scale proposed detector, with similar innovation needed to lower the cost per channel for an array like GRAND, with a full size of $2$x$10^5$ channels. For a forest detector, the trees are simply connected to readout electronics, obviating the need to design, deploy, and maintain manufactured antennas.  

{\bf Challenges and opportunities.---} Detailed experimental study of the efficiency of trees as antennas in specific forests is beyond the scope of this initial letter. Further studies of the efficiency of trees as antennas in the VHF range, where the background noise from the atmosphere and galaxy are lower (and where existing $\nu_{\tau}$ detectors operate), as well as studies of the effect of polarization, gain pattern, and phasing, are required to establish the validity of a forest detector. Also important to study is the uniformity of trees that comprise an array. These are all needed prior to generating a full projected detector sensitivity. A design challenge unique to a forest detector could be providing power to the readout electronics. In a dense forest, finding sky for a solar array may be challenging, requiring a tall mast, or even unfeasible. We  do not quantify the comparative difficulty of this versus the instrumentation of a full array with manufactured antennas. Finally, a study of background radio sources unique to a forested environment is needed. 

Such a detector could also be useful for cosmic ray studies. And, built responsibly, will have benefits for forest conservation, particularly in places where forests are multiple use. A forest detector could also motivate the large-scale reforesting of land, to grow a neutrino detector for future generations.

{\bf Conclusion.---} The forest may be an efficient $\nu_{\tau}$ detector. Using trees as the antennas for a full scale array has numerous benefits, including the complete removal of the need for designing and deploying antennas. The trees are simply instrumented. The author urges that, should this idea be tested or implemented, the experimenters take the utmost care in protecting and maintaining the forest they instrument, and not do damage to this precious and fragile resource shared by all the creatures of the Earth. Such a detector must be built in harmony with, and with respect for, nature; otherwise, this idea is not worth trying.

{\bf Acknowledgements.---}
The author thanks J.~F.~Beacom, D.~Besson, I.~Esteban, S.~Wissel, and especially C.~Deaconu and K.~D.~de Vries for helpful feedback and thoughtful suggestions. The author recognizes support from the John D. and Catherine T. MacArthur Foundation, and the National Science Foundation under award No. 2306424. 

\bibliography{main}

\begin{thebibliography}{36}%
\makeatletter
\providecommand \@ifxundefined [1]{%
 \@ifx{#1\undefined}
}%
\providecommand \@ifnum [1]{%
 \ifnum #1\expandafter \@firstoftwo
 \else \expandafter \@secondoftwo
 \fi
}%
\providecommand \@ifx [1]{%
 \ifx #1\expandafter \@firstoftwo
 \else \expandafter \@secondoftwo
 \fi
}%
\providecommand \natexlab [1]{#1}%
\providecommand \enquote  [1]{``#1''}%
\providecommand \bibnamefont  [1]{#1}%
\providecommand \bibfnamefont [1]{#1}%
\providecommand \citenamefont [1]{#1}%
\providecommand \href@noop [0]{\@secondoftwo}%
\providecommand \href [0]{\begingroup \@sanitize@url \@href}%
\providecommand \@href[1]{\@@startlink{#1}\@@href}%
\providecommand \@@href[1]{\endgroup#1\@@endlink}%
\providecommand \@sanitize@url [0]{\catcode `\\12\catcode `\$12\catcode `\&12\catcode `\#12\catcode `\^12\catcode `\_12\catcode `\%12\relax}%
\providecommand \@@startlink[1]{}%
\providecommand \@@endlink[0]{}%
\providecommand \url  [0]{\begingroup\@sanitize@url \@url }%
\providecommand \@url [1]{\endgroup\@href {#1}{\urlprefix }}%
\providecommand \urlprefix  [0]{URL }%
\providecommand \Eprint [0]{\href }%
\providecommand \doibase [0]{http://dx.doi.org/}%
\providecommand \selectlanguage [0]{\@gobble}%
\providecommand \bibinfo  [0]{\@secondoftwo}%
\providecommand \bibfield  [0]{\@secondoftwo}%
\providecommand \translation [1]{[#1]}%
\providecommand \BibitemOpen [0]{}%
\providecommand \bibitemStop [0]{}%
\providecommand \bibitemNoStop [0]{.\EOS\space}%
\providecommand \EOS [0]{\spacefactor3000\relax}%
\providecommand \BibitemShut  [1]{\csname bibitem#1\endcsname}%
\let\auto@bib@innerbib\@empty
\bibitem [{\citenamefont {Ahrens}\ \emph {et~al.}(2003)\citenamefont {Ahrens} \emph {et~al.}}]{IceCube:2002eys}%
  \BibitemOpen
  \bibfield  {author} {\bibinfo {author} {\bibfnamefont {J.}~\bibnamefont {Ahrens}} \emph {et~al.} (\bibinfo {collaboration} {IceCube}),\ }\href {\doibase 10.1016/S0920-5632(03)01337-9} {\bibfield  {journal} {\bibinfo  {journal} {Nucl. Phys. B Proc. Suppl.}\ }\textbf {\bibinfo {volume} {118}},\ \bibinfo {pages} {388} (\bibinfo {year} {2003})},\ \Eprint {http://arxiv.org/abs/astro-ph/0209556} {arXiv:astro-ph/0209556} \BibitemShut {NoStop}%
\bibitem [{\citenamefont {Belolaptikov}\ \emph {et~al.}(1997)\citenamefont {Belolaptikov} \emph {et~al.}}]{BAIKAL:1997iok}%
  \BibitemOpen
  \bibfield  {author} {\bibinfo {author} {\bibfnamefont {I.~A.}\ \bibnamefont {Belolaptikov}} \emph {et~al.} (\bibinfo {collaboration} {BAIKAL}),\ }\href {\doibase 10.1016/S0927-6505(97)00022-4} {\bibfield  {journal} {\bibinfo  {journal} {Astropart. Phys.}\ }\textbf {\bibinfo {volume} {7}},\ \bibinfo {pages} {263} (\bibinfo {year} {1997})}\BibitemShut {NoStop}%
\bibitem [{\citenamefont {Adrian-Martinez}\ \emph {et~al.}(2016)\citenamefont {Adrian-Martinez} \emph {et~al.}}]{KM3Net:2016zxf}%
  \BibitemOpen
  \bibfield  {author} {\bibinfo {author} {\bibfnamefont {S.}~\bibnamefont {Adrian-Martinez}} \emph {et~al.} (\bibinfo {collaboration} {KM3Net}),\ }\href {\doibase 10.1088/0954-3899/43/8/084001} {\bibfield  {journal} {\bibinfo  {journal} {J. Phys. G}\ }\textbf {\bibinfo {volume} {43}},\ \bibinfo {pages} {084001} (\bibinfo {year} {2016})},\ \Eprint {http://arxiv.org/abs/1601.07459} {arXiv:1601.07459 [astro-ph.IM]} \BibitemShut {NoStop}%
\bibitem [{\citenamefont {Allison}\ \emph {et~al.}(2015)\citenamefont {Allison} \emph {et~al.}}]{ARA:2014fyf}%
  \BibitemOpen
  \bibfield  {author} {\bibinfo {author} {\bibfnamefont {P.}~\bibnamefont {Allison}} \emph {et~al.} (\bibinfo {collaboration} {ARA}),\ }\href {\doibase 10.1016/j.astropartphys.2015.04.006} {\bibfield  {journal} {\bibinfo  {journal} {Astropart. Phys.}\ }\textbf {\bibinfo {volume} {70}},\ \bibinfo {pages} {62} (\bibinfo {year} {2015})},\ \Eprint {http://arxiv.org/abs/1404.5285} {arXiv:1404.5285 [astro-ph.HE]} \BibitemShut {NoStop}%
\bibitem [{\citenamefont {Gorham}\ \emph {et~al.}(2009)\citenamefont {Gorham} \emph {et~al.}}]{ANITA:2008mzi}%
  \BibitemOpen
  \bibfield  {author} {\bibinfo {author} {\bibfnamefont {P.~W.}\ \bibnamefont {Gorham}} \emph {et~al.} (\bibinfo {collaboration} {ANITA}),\ }\href {\doibase 10.1016/j.astropartphys.2009.05.003} {\bibfield  {journal} {\bibinfo  {journal} {Astropart. Phys.}\ }\textbf {\bibinfo {volume} {32}},\ \bibinfo {pages} {10} (\bibinfo {year} {2009})},\ \Eprint {http://arxiv.org/abs/0812.1920} {arXiv:0812.1920 [astro-ph]} \BibitemShut {NoStop}%
\bibitem [{\citenamefont {Barwick}\ \emph {et~al.}(2015)\citenamefont {Barwick} \emph {et~al.}}]{ARIANNA:2014fsk}%
  \BibitemOpen
  \bibfield  {author} {\bibinfo {author} {\bibfnamefont {S.~W.}\ \bibnamefont {Barwick}} \emph {et~al.} (\bibinfo {collaboration} {ARIANNA}),\ }\href {\doibase 10.1016/j.astropartphys.2015.04.002} {\bibfield  {journal} {\bibinfo  {journal} {Astropart. Phys.}\ }\textbf {\bibinfo {volume} {70}},\ \bibinfo {pages} {12} (\bibinfo {year} {2015})},\ \Eprint {http://arxiv.org/abs/1410.7352} {arXiv:1410.7352 [astro-ph.HE]} \BibitemShut {NoStop}%
\bibitem [{\citenamefont {Nam}\ \emph {et~al.}(2016)\citenamefont {Nam} \emph {et~al.}}]{Nam:2016cib}%
  \BibitemOpen
  \bibfield  {author} {\bibinfo {author} {\bibfnamefont {J.~W.}\ \bibnamefont {Nam}} \emph {et~al.},\ }\href {\doibase 10.1142/S0218271816450139} {\bibfield  {journal} {\bibinfo  {journal} {Int. J. Mod. Phys. D}\ }\textbf {\bibinfo {volume} {25}},\ \bibinfo {pages} {1645013} (\bibinfo {year} {2016})}\BibitemShut {NoStop}%
\bibitem [{\citenamefont {\'Alvarez-Mu\~niz}\ \emph {et~al.}(2020)\citenamefont {\'Alvarez-Mu\~niz} \emph {et~al.}}]{GRAND:2018iaj}%
  \BibitemOpen
  \bibfield  {author} {\bibinfo {author} {\bibfnamefont {J.}~\bibnamefont {\'Alvarez-Mu\~niz}} \emph {et~al.} (\bibinfo {collaboration} {GRAND}),\ }\href {\doibase 10.1007/s11433-018-9385-7} {\bibfield  {journal} {\bibinfo  {journal} {Sci. China Phys. Mech. Astron.}\ }\textbf {\bibinfo {volume} {63}},\ \bibinfo {pages} {219501} (\bibinfo {year} {2020})},\ \Eprint {http://arxiv.org/abs/1810.09994} {arXiv:1810.09994 [astro-ph.HE]} \BibitemShut {NoStop}%
\bibitem [{\citenamefont {Otte}(2019)}]{Otte:2018uxj}%
  \BibitemOpen
  \bibfield  {author} {\bibinfo {author} {\bibfnamefont {A.~N.}\ \bibnamefont {Otte}},\ }\href {\doibase 10.1103/PhysRevD.99.083012} {\bibfield  {journal} {\bibinfo  {journal} {Phys. Rev. D}\ }\textbf {\bibinfo {volume} {99}},\ \bibinfo {pages} {083012} (\bibinfo {year} {2019})},\ \Eprint {http://arxiv.org/abs/1811.09287} {arXiv:1811.09287 [astro-ph.IM]} \BibitemShut {NoStop}%
\bibitem [{\citenamefont {Aguilar}\ \emph {et~al.}(2021)\citenamefont {Aguilar} \emph {et~al.}}]{RNO-G:2020rmc}%
  \BibitemOpen
  \bibfield  {author} {\bibinfo {author} {\bibfnamefont {J.~A.}\ \bibnamefont {Aguilar}} \emph {et~al.} (\bibinfo {collaboration} {RNO-G}),\ }\href {\doibase 10.1088/1748-0221/16/03/P03025} {\bibfield  {journal} {\bibinfo  {journal} {JINST}\ }\textbf {\bibinfo {volume} {16}},\ \bibinfo {pages} {P03025} (\bibinfo {year} {2021})},\ \Eprint {http://arxiv.org/abs/2010.12279} {arXiv:2010.12279 [astro-ph.IM]} \BibitemShut {NoStop}%
\bibitem [{\citenamefont {Abarr}\ \emph {et~al.}(2021)\citenamefont {Abarr} \emph {et~al.}}]{PUEO:2020bnn}%
  \BibitemOpen
  \bibfield  {author} {\bibinfo {author} {\bibfnamefont {Q.}~\bibnamefont {Abarr}} \emph {et~al.} (\bibinfo {collaboration} {PUEO}),\ }\href {\doibase 10.1088/1748-0221/16/08/P08035} {\bibfield  {journal} {\bibinfo  {journal} {JINST}\ }\textbf {\bibinfo {volume} {16}},\ \bibinfo {pages} {P08035} (\bibinfo {year} {2021})},\ \Eprint {http://arxiv.org/abs/2010.02892} {arXiv:2010.02892 [astro-ph.IM]} \BibitemShut {NoStop}%
\bibitem [{\citenamefont {Olinto}\ \emph {et~al.}(2021)\citenamefont {Olinto} \emph {et~al.}}]{POEMMA:2020ykm}%
  \BibitemOpen
  \bibfield  {author} {\bibinfo {author} {\bibfnamefont {A.~V.}\ \bibnamefont {Olinto}} \emph {et~al.} (\bibinfo {collaboration} {POEMMA}),\ }\href {\doibase 10.1088/1475-7516/2021/06/007} {\bibfield  {journal} {\bibinfo  {journal} {JCAP}\ }\textbf {\bibinfo {volume} {06}},\ \bibinfo {pages} {007} (\bibinfo {year} {2021})},\ \Eprint {http://arxiv.org/abs/2012.07945} {arXiv:2012.07945 [astro-ph.IM]} \BibitemShut {NoStop}%
\bibitem [{\citenamefont {Wissel}\ \emph {et~al.}(2020)\citenamefont {Wissel} \emph {et~al.}}]{Wissel:2020sec}%
  \BibitemOpen
  \bibfield  {author} {\bibinfo {author} {\bibfnamefont {S.}~\bibnamefont {Wissel}} \emph {et~al.},\ }\href {\doibase 10.1088/1475-7516/2020/11/065} {\bibfield  {journal} {\bibinfo  {journal} {JCAP}\ }\textbf {\bibinfo {volume} {11}},\ \bibinfo {pages} {065} (\bibinfo {year} {2020})},\ \Eprint {http://arxiv.org/abs/2004.12718} {arXiv:2004.12718 [astro-ph.IM]} \BibitemShut {NoStop}%
\bibitem [{\citenamefont {Romero-Wolf}\ \emph {et~al.}(2020)\citenamefont {Romero-Wolf} \emph {et~al.}}]{Romero-Wolf:2020pzh}%
  \BibitemOpen
  \bibfield  {author} {\bibinfo {author} {\bibfnamefont {A.}~\bibnamefont {Romero-Wolf}} \emph {et~al.},\ }in\ \href@noop {} {\emph {\bibinfo {booktitle} {{Latin American Strategy Forum for Research Infrastructure}}}}\ (\bibinfo {year} {2020})\ \Eprint {http://arxiv.org/abs/2002.06475} {arXiv:2002.06475 [astro-ph.IM]} \BibitemShut {NoStop}%
\bibitem [{\citenamefont {Agostini}\ \emph {et~al.}(2020)\citenamefont {Agostini} \emph {et~al.}}]{P-ONE:2020ljt}%
  \BibitemOpen
  \bibfield  {author} {\bibinfo {author} {\bibfnamefont {M.}~\bibnamefont {Agostini}} \emph {et~al.} (\bibinfo {collaboration} {P-ONE}),\ }\href {\doibase 10.1038/s41550-020-1182-4} {\bibfield  {journal} {\bibinfo  {journal} {Nature Astron.}\ }\textbf {\bibinfo {volume} {4}},\ \bibinfo {pages} {913} (\bibinfo {year} {2020})},\ \Eprint {http://arxiv.org/abs/2005.09493} {arXiv:2005.09493 [astro-ph.HE]} \BibitemShut {NoStop}%
\bibitem [{\citenamefont {Prohira}\ \emph {et~al.}(2021)\citenamefont {Prohira} \emph {et~al.}}]{RadarEchoTelescope:2021rca}%
  \BibitemOpen
  \bibfield  {author} {\bibinfo {author} {\bibfnamefont {S.}~\bibnamefont {Prohira}} \emph {et~al.} (\bibinfo {collaboration} {Radar Echo Telescope}),\ }\href {\doibase 10.1103/PhysRevD.104.102006} {\bibfield  {journal} {\bibinfo  {journal} {Phys. Rev. D}\ }\textbf {\bibinfo {volume} {104}},\ \bibinfo {pages} {102006} (\bibinfo {year} {2021})},\ \Eprint {http://arxiv.org/abs/2104.00459} {arXiv:2104.00459 [astro-ph.IM]} \BibitemShut {NoStop}%
\bibitem [{\citenamefont {Ackermann}\ \emph {et~al.}(2022)\citenamefont {Ackermann} \emph {et~al.}}]{Ackermann:2022rqc}%
  \BibitemOpen
  \bibfield  {author} {\bibinfo {author} {\bibfnamefont {M.}~\bibnamefont {Ackermann}} \emph {et~al.},\ }\href@noop {} {\  (\bibinfo {year} {2022})},\ \Eprint {http://arxiv.org/abs/2203.08096} {arXiv:2203.08096 [hep-ph]} \BibitemShut {NoStop}%
\bibitem [{\citenamefont {Southall}\ \emph {et~al.}(2023)\citenamefont {Southall} \emph {et~al.}}]{Southall:2022yil}%
  \BibitemOpen
  \bibfield  {author} {\bibinfo {author} {\bibfnamefont {D.}~\bibnamefont {Southall}} \emph {et~al.},\ }\href {\doibase 10.1016/j.nima.2022.167889} {\bibfield  {journal} {\bibinfo  {journal} {Nucl. Instrum. Meth. A}\ }\textbf {\bibinfo {volume} {1048}},\ \bibinfo {pages} {167889} (\bibinfo {year} {2023})},\ \Eprint {http://arxiv.org/abs/2206.09660} {arXiv:2206.09660 [astro-ph.IM]} \BibitemShut {NoStop}%
\bibitem [{\citenamefont {Martineau-Huynh}(2019)}]{Martineau-Huynh:2019bgk}%
  \BibitemOpen
  \bibfield  {author} {\bibinfo {author} {\bibfnamefont {O.}~\bibnamefont {Martineau-Huynh}} (\bibinfo {collaboration} {GRAND}),\ }\href {\doibase 10.1051/epjconf/201921006007} {\bibfield  {journal} {\bibinfo  {journal} {EPJ Web Conf.}\ }\textbf {\bibinfo {volume} {210}},\ \bibinfo {pages} {06007} (\bibinfo {year} {2019})},\ \Eprint {http://arxiv.org/abs/1903.04803} {arXiv:1903.04803 [astro-ph.IM]} \BibitemShut {NoStop}%
\bibitem [{\citenamefont {Otte}\ \emph {et~al.}(2023)\citenamefont {Otte} \emph {et~al.}}]{Otte:2023osf}%
  \BibitemOpen
  \bibfield  {author} {\bibinfo {author} {\bibfnamefont {A.~N.}\ \bibnamefont {Otte}} \emph {et~al.},\ }\href {\doibase 10.22323/1.444.1170} {\bibfield  {journal} {\bibinfo  {journal} {PoS}\ }\textbf {\bibinfo {volume} {ICRC2023}},\ \bibinfo {pages} {1170} (\bibinfo {year} {2023})}\BibitemShut {NoStop}%
\bibitem [{\citenamefont {Wang}\ \emph {et~al.}(2022)\citenamefont {Wang} \emph {et~al.}}]{TAROGE:2022soh}%
  \BibitemOpen
  \bibfield  {author} {\bibinfo {author} {\bibfnamefont {S.-H.}\ \bibnamefont {Wang}} \emph {et~al.} (\bibinfo {collaboration} {TAROGE, Arianna}),\ }\href {\doibase 10.1088/1475-7516/2022/11/022} {\bibfield  {journal} {\bibinfo  {journal} {JCAP}\ }\textbf {\bibinfo {volume} {11}},\ \bibinfo {pages} {022} (\bibinfo {year} {2022})},\ \Eprint {http://arxiv.org/abs/2207.10616} {arXiv:2207.10616 [astro-ph.HE]} \BibitemShut {NoStop}%
\bibitem [{\citenamefont {{Squier, George O.}}(1919)}]{treeAntenna}%
  \BibitemOpen
  \bibfield  {author} {\bibinfo {author} {\bibnamefont {{Squier, George O.}}},\ }\href@noop {} {\bibfield  {journal} {\bibinfo  {journal} {{Scientific American}}\ ,\ \bibinfo {pages} {{624}}} (\bibinfo {year} {{1919}})}\BibitemShut {NoStop}%
\bibitem [{\citenamefont {Ikrath}\ \emph {et~al.}(1975)\citenamefont {Ikrath}, \citenamefont {Kennebeck},\ and\ \citenamefont {Hoverter}}]{ikrath1975trees}%
  \BibitemOpen
  \bibfield  {author} {\bibinfo {author} {\bibfnamefont {K.}~\bibnamefont {Ikrath}}, \bibinfo {author} {\bibfnamefont {W.}~\bibnamefont {Kennebeck}}, \ and\ \bibinfo {author} {\bibfnamefont {R.}~\bibnamefont {Hoverter}},\ }\href@noop {} {\bibfield  {journal} {\bibinfo  {journal} {IEEE Transactions on Antennas and Propagation}\ }\textbf {\bibinfo {volume} {23}},\ \bibinfo {pages} {137} (\bibinfo {year} {1975})}\BibitemShut {NoStop}%
\bibitem [{\citenamefont {Skirvseth}(1971)}]{ad735330}%
  \BibitemOpen
  \bibfield  {author} {\bibinfo {author} {\bibfnamefont {K.}~\bibnamefont {Skirvseth}},\ }\href@noop {} {\bibfield  {journal} {\bibinfo  {journal} {Army Electronics Command Fort Monmouth NJ}\ } (\bibinfo {year} {1971})}\BibitemShut {NoStop}%
\bibitem [{\citenamefont {IKRATH}\ \emph {et~al.}(1973)\citenamefont {IKRATH}, \citenamefont {MURPHY},\ and\ \citenamefont {KENNEBECK}}]{ikrath1973utilization}%
  \BibitemOpen
  \bibfield  {author} {\bibinfo {author} {\bibfnamefont {K.}~\bibnamefont {IKRATH}}, \bibinfo {author} {\bibfnamefont {K.}~\bibnamefont {MURPHY}}, \ and\ \bibinfo {author} {\bibfnamefont {W.}~\bibnamefont {KENNEBECK}},\ }\href@noop {} {\  (\bibinfo {year} {1973})}\BibitemShut {NoStop}%
\bibitem [{\citenamefont {Ikrath}\ \emph {et~al.}(1972)\citenamefont {Ikrath}, \citenamefont {Kennebeck}, \citenamefont {Hoverter},\ and\ \citenamefont {NJ}}]{ikrath1972performance}%
  \BibitemOpen
  \bibfield  {author} {\bibinfo {author} {\bibfnamefont {K.}~\bibnamefont {Ikrath}}, \bibinfo {author} {\bibfnamefont {W.}~\bibnamefont {Kennebeck}}, \bibinfo {author} {\bibfnamefont {R.~T.}\ \bibnamefont {Hoverter}}, \ and\ \bibinfo {author} {\bibfnamefont {A.~E. C. F.~M.}\ \bibnamefont {NJ}},\ }\href@noop {} {\emph {\bibinfo {title} {Performance of Trees as Radio Antennas in Tropical Jungle Forests:(Panama Canal Zone Experiments)}}}\ (\bibinfo  {publisher} {US Army Electronics Command},\ \bibinfo {year} {1972})\BibitemShut {NoStop}%
\bibitem [{\citenamefont {Kar}\ \emph {et~al.}(2010)\citenamefont {Kar}, \citenamefont {Chakrabarty},\ and\ \citenamefont {Sarkar}}]{kar2010trees}%
  \BibitemOpen
  \bibfield  {author} {\bibinfo {author} {\bibfnamefont {S.~J.}\ \bibnamefont {Kar}}, \bibinfo {author} {\bibfnamefont {A.}~\bibnamefont {Chakrabarty}}, \ and\ \bibinfo {author} {\bibfnamefont {B.}~\bibnamefont {Sarkar}},\ }in\ \href@noop {} {\emph {\bibinfo {booktitle} {2010 Annual IEEE India Conference (INDICON)}}}\ (\bibinfo {organization} {IEEE},\ \bibinfo {year} {2010})\ pp.\ \bibinfo {pages} {1--3}\BibitemShut {NoStop}%
\bibitem [{\citenamefont {Liao}\ \emph {et~al.}(2023)\citenamefont {Liao}, \citenamefont {Killough},\ and\ \citenamefont {Kerekes}}]{liao2023performance}%
  \BibitemOpen
  \bibfield  {author} {\bibinfo {author} {\bibfnamefont {D.}~\bibnamefont {Liao}}, \bibinfo {author} {\bibfnamefont {S.~M.}\ \bibnamefont {Killough}}, \ and\ \bibinfo {author} {\bibfnamefont {R.~A.}\ \bibnamefont {Kerekes}},\ }\href@noop {} {\bibfield  {journal} {\bibinfo  {journal} {IEEE Access}\ }\textbf {\bibinfo {volume} {11}},\ \bibinfo {pages} {136041} (\bibinfo {year} {2023})}\BibitemShut {NoStop}%
\bibitem [{\citenamefont {Tam}\ and\ \citenamefont {Rockway}()}]{treeantennapatent}%
  \BibitemOpen
  \bibfield  {author} {\bibinfo {author} {\bibfnamefont {D.~W.~S.}\ \bibnamefont {Tam}}\ and\ \bibinfo {author} {\bibfnamefont {J.~W.}\ \bibnamefont {Rockway}},\ }\href@noop {} {\enquote {\bibinfo {title} {Multi-band tree antenna},}\ }\bibinfo {note} {Patent No. US8094083B1}\BibitemShut {NoStop}%
\bibitem [{\citenamefont {Wait}\ \emph {et~al.}(1974)\citenamefont {Wait}, \citenamefont {Ott}, \citenamefont {Telfer},\ and\ \citenamefont {OFFICE}}]{wait1974workshop}%
  \BibitemOpen
  \bibfield  {author} {\bibinfo {author} {\bibfnamefont {J.}~\bibnamefont {Wait}}, \bibinfo {author} {\bibfnamefont {R.}~\bibnamefont {Ott}}, \bibinfo {author} {\bibfnamefont {T.}~\bibnamefont {Telfer}}, \ and\ \bibinfo {author} {\bibfnamefont {A.~C. C. F. H. A. A.~C.}\ \bibnamefont {OFFICE}},\ }\href@noop {} {\bibfield  {journal} {\bibinfo  {journal} {US Army Communications Command, Fort Huachuca, AZ, Tech. Rep. No. ACC-ACO-1-74}\ } (\bibinfo {year} {1974})}\BibitemShut {NoStop}%
\bibitem [{\citenamefont {Sturgill}(1967)}]{sturgill1967tropical}%
  \BibitemOpen
  \bibfield  {author} {\bibinfo {author} {\bibfnamefont {L.~G.}\ \bibnamefont {Sturgill}},\ }\href@noop {} {\bibfield  {journal} {\bibinfo  {journal} {Research and Engineering Dept., Atlantic Research Corp., Alexandria, Va}\ }\textbf {\bibinfo {volume} {1}} (\bibinfo {year} {1967})}\BibitemShut {NoStop}%
\bibitem [{\citenamefont {Hicks}\ \emph {et~al.}(1969)\citenamefont {Hicks}, \citenamefont {Murphy}, \citenamefont {Patrick},\ and\ \citenamefont {Sturgill}}]{hicks1969tropical}%
  \BibitemOpen
  \bibfield  {author} {\bibinfo {author} {\bibfnamefont {J.~J.}\ \bibnamefont {Hicks}}, \bibinfo {author} {\bibfnamefont {A.}~\bibnamefont {Murphy}}, \bibinfo {author} {\bibfnamefont {E.}~\bibnamefont {Patrick}}, \ and\ \bibinfo {author} {\bibfnamefont {L.}~\bibnamefont {Sturgill}},\ }\href@noop {} {\bibfield  {journal} {\bibinfo  {journal} {Research and Engineering Dept., Atlantic Research Corp., Alexandria, Va}\ }\textbf {\bibinfo {volume} {2}} (\bibinfo {year} {1969})}\BibitemShut {NoStop}%
\bibitem [{\citenamefont {Robertson}\ \emph {et~al.}(1969)\citenamefont {Robertson}, \citenamefont {Hicks}, \citenamefont {Sykes},\ and\ \citenamefont {Anti}}]{robertson1969tropical}%
  \BibitemOpen
  \bibfield  {author} {\bibinfo {author} {\bibfnamefont {R.~G.}\ \bibnamefont {Robertson}}, \bibinfo {author} {\bibfnamefont {J.~J.}\ \bibnamefont {Hicks}}, \bibinfo {author} {\bibfnamefont {C.~B.}\ \bibnamefont {Sykes}}, \ and\ \bibinfo {author} {\bibfnamefont {P.~A.}\ \bibnamefont {Anti}},\ }\href@noop {} {\bibfield  {journal} {\bibinfo  {journal} {Research and Engineering Dept., Atlantic Research Corp., Alexandria, Va}\ }\textbf {\bibinfo {volume} {3}} (\bibinfo {year} {1969})}\BibitemShut {NoStop}%
\bibitem [{\citenamefont {Nelson}\ and\ \citenamefont {OH}(1980)}]{nelson1980uhf}%
  \BibitemOpen
  \bibfield  {author} {\bibinfo {author} {\bibfnamefont {R.~A.}\ \bibnamefont {Nelson}}\ and\ \bibinfo {author} {\bibfnamefont {B.~C.~D.}\ \bibnamefont {OH}},\ }\href@noop {} {\emph {\bibinfo {title} {UHF (Ultra-High-Frequency) Propagation in Vegetative Media.}}},\ \bibinfo {type} {Tech. Rep.}\ (\bibinfo {year} {1980})\BibitemShut {NoStop}%
\bibitem [{\citenamefont {Richardson}(2000)}]{richardson2000ecology}%
  \BibitemOpen
  \bibfield  {author} {\bibinfo {author} {\bibfnamefont {D.~M.}\ \bibnamefont {Richardson}},\ }\href@noop {} {\emph {\bibinfo {title} {Ecology and biogeography of Pinus}}}\ (\bibinfo  {publisher} {Cambridge University Press},\ \bibinfo {year} {2000})\BibitemShut {NoStop}%
\bibitem [{\citenamefont {Prohira}(2024)}]{tree_exp_prelim}%
  \BibitemOpen
  \bibfield  {author} {\bibinfo {author} {\bibfnamefont {S.}~\bibnamefont {Prohira}},\ }\href@noop {} {\bibfield  {journal} {\bibinfo  {journal} {In Preparation}\ } (\bibinfo {year} {2024})}\BibitemShut {NoStop}%
\end{thebibliography}%
\end{document}